\documentclass[final,5p,times,twocolumn,sort&compress]{elsarticle}

\usepackage{amssymb}
\usepackage{amsthm}
\usepackage{natbib}

\usepackage{multirow}
\usepackage{amsmath}

\begin{document}

\begin{frontmatter}



\title{On the ground-state degeneracy and entropy in a double-tetrahedral chain \\ formed  by the localized Ising spins and mobile electrons}


\author{Lucia G\'alisov\'a}
\ead{galisova.lucia@gmail.com}
\address[label1]{Department of Applied Mathematics and Informatics,
				Faculty of Mechanical Engineering, Technical University of Ko\v{s}ice,
	            Letn\'a 9, 042 00 Ko\v{s}ice, Slovakia}
\begin{abstract}
Ground-state properties of a hybrid double-tetrahedral chain, in which the localized Ising spins regularly alternate with triangular plaquettes occupied by a variable number of mobile electrons, are exactly investigated. We demonstrate that the zero-temperature phase diagram of the model involves several non-degenerate, two-fold degenerate and macroscopically degenerate chiral phases. Low-temperature dependencies of the entropy and specific heat are also examined in order to gain a deeper insight into the degeneracy of individual ground-state phases and phase transitions. It is shown that a diversity of the ground-state degeneracy manifests itself in multiple-peak structures of both thermodynamic quantities. A remarkable temperature dependencies of the specific heat with two and three Schottky-type maxima are discussed in detail.
\end{abstract}


\end{frontmatter}


\section{Introduction}
\label{intro}

Low-dimensional correlated spin-electron systems, which are rigorously treatable by means of the standard transfer-matrix method~\cite{Bax82} and/or the generalized decoration-iteration mapping transformation~\cite{Syo72}, represent an excellent play-ground for theoretical investigation of many unusual cooperative phenomena. Despite of their simplicity, these models may provide novel insight into macroscopic degeneracy of the ground state~\cite{Per08, Per09, Nal14, Cis14, Gal15a, Gal15b, Str16, Gal17}, rational~\cite{Per08, Nal14, Gal15a, Gal15b, Gal17} and irrational (doping-dependent)~\cite{Str16} plateaus in low-temperature magnetization curves, entanglement properties~\cite{Nal14}, double- and triple-peak structure in temperature dependencies of the specific heat~\cite{Per08, Nal14, Gal15a, Cis14}, as well as enhanced magnetocaloric effect~\cite{Per09, Gal15a, Gal15b, Str16, Gal17}. Moreover, hybrid spin-electron systems with the geometry of decorated planar lattices quite well demonstrate the presence of the spontaneous long-range order in the ground state~\cite{Gal11} and also uncommon reentrant phase transitions at non-zero temperatures~\cite{Dor14}.

\begin{figure}[t!]
\vspace{0mm}
\centering
  \includegraphics[width=0.4\textwidth]{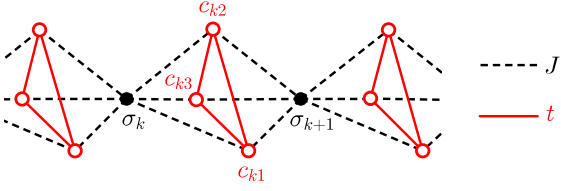}
\caption{(Color online) A part of the spin-electron double-tetrahedral chain. Full circles denote nodal sites occupied by the localized Ising spins $\sigma = 1/2$ and empty circles forming triangular plaquettes are available to mobile electrons.}
\label{fig:1}
\end{figure}

Motivated by aforementioned facts, in the present paper we investigate an exactly solvable double-tetrahedral chain, in which nodal lattice sites occupied by the localized Ising spins regularly alternate with triangular clusters with the dynamics described by Hubbard model (see Fig.~\ref{fig:1}). As recently shown~\cite{Gal15a, Gal15b, Gal17}, this hybrid model provides an excellent prototype, which allows a rigorous study of the macroscopic degeneracy at absolute zero temperature, the associated residual entropy, as well as other thermodynamic quantities. Another important stimulus for investigating the aforementioned model relates to the copper-based polymeric compound Cu$_3$Mo$_2$O$_9$~\cite{Has08, Mat12}, which represents the experimental realizations of the double-tetrahedral chain structure.

\section{Model and its exact solution}
\label{sec:2}

To be specific, we consider a spin-electron double-tetrahedral chain defined through the Hamiltonian:
\begin{eqnarray}
\label{eq:H_tot}
\hspace{-3mm}
\hat{{\cal H}}\!\!\!\!&=&\!\!\!\! \sum_{k=1}^N \hat{{\cal H}}_k,
\\
\hspace{-3mm}
\label{eq:Hk}
\hat{{\cal H}}_k\!\!\!\!&=&\!\!\!\!
-\,t\sum_{j = 1}^3\sum_{\gamma\in\{\uparrow,\downarrow\}}\left(\hat{c}_{kj,\gamma}^{\dagger}\hat{c}_{k(j+1)_{\rm mod\,3},\gamma} + \hat{c}_{k(j+1)_{\rm mod\,3},\gamma}^{\dagger}\hat{c}_{kj,\gamma}\right)
\nonumber \\
\hspace{-3mm}
\!\!\!\!&&\!\!\!\!
-\, J\hat{S}_k^z(\sigma_k^z + \sigma_{k+1}^z) - \mu\hat{n}_k - H_{e}\hat{S}_k^z - \frac{H_{I}}{2}(\sigma_k^z + \sigma_{k+1}^z).
\end{eqnarray}
In above, $N$ represents the total number of nodal lattice sites occupied by the localized Ising spins $\sigma=1/2$ ($N\to\infty$), $\hat{c}_{kj,\gamma}^{\dagger}$ and $\hat{c}_{kj,\gamma}$ mean fermionic creation and annihilation operators for mobile electrons delocalized over $k$th triangular cluster with the spin $\gamma\in \{\uparrow,\downarrow\}$, respectively, $\hat{S}_k^z = \sum_{j = 1}^{3}(\hat{n}_{kj,\uparrow}-\hat{n}_{kj,\downarrow})/2$ labels the $z$th component of the total spin operator, $\hat{n}_{k} = \sum_{j = 1}^{3}(\hat{n}_{kj,\uparrow}+\hat{n}_{kj,\downarrow})$ denotes the total number operator corresponding to electrons in $k$th triangular cluster ($\hat{n}_{kj,\gamma} = \hat{c}_{kj,\gamma}^{\dag}\hat{c}_{kj,\gamma}$) and $\sigma_{k}^{z}= \pm 1/2$ marks the Ising spin occupied the $k$th nodal lattice site. The term $t>0$ represents hopping parameter, which takes into account the kinetic energy of mobile electrons, $J$ stands for the exchange interaction between mobile electrons and their nearest Ising neighbors and $\mu$ is the chemical potential allowing one to tune the total number of electrons in the system. The last two terms in Eq.~(\ref{eq:Hk}) represent Zeeman energies of mobile electrons and Ising spins in a presence of the 'effective' magnetic fields $H_{e} = g_e\mu_{\rm B}B$ and $H_{I} = g_I\mu_{\rm B}B$ ($g_e$ and $g_I$ are Land\'e g-factors of electrons and Ising spins, respectively, $\mu_{\rm B}$ is Bohr magneton). Finally, the modulo 3 operation is introduced into Eq.~(\ref{eq:Hk}) to ensure the periodic boundary condition $\hat{c}_{k4,\gamma}^{\dagger}=\hat{c}_{k1,\gamma}^{\dagger}$ ($\hat{c}_{k4,\gamma}=\hat{c}_{k1,\gamma}$) for the three-site electron subsystem.

The cluster Hamiltonian~(\ref{eq:Hk}) can be written as $64\times64$ matrix if a number of mobile electrons at each triangular plaquette varies from zero up to six. Thanks to a validity of the commutation relations $[\hat{{\cal H}}_k, \hat{n}_{k}] = 0$ and $[\hat{{\cal H}}_k, \hat{S}_k^z] = 0$, the matrix representation of the Hamiltonian~(\ref{eq:Hk}) has the block-diagonal form. Individual matrix blocks correspond to the orthogonal Hilbert subspaces with different but fixed numbers $n_k$ of mobile electrons and distinct values of the total spins $S_k^z$. Diagonalization of these blocks gives a full spectrum of eigenvalues for the cluster Hamiltonian~(\ref{eq:Hk}), which can be further employed both for a comprehensive analysis of the ground-state configuration and an analytical calculation of the grand-canonical potential of the model:
\begin{eqnarray}
\label{eq:Omega}
\Omega\!\!\!\!&=&\!\!\!\!
-k_{\rm B}T\!\lim_{N\to\infty}\frac{1}{N}\ln\Bigg[\!\sum_{\{\sigma_k\!\}}\mathrm{Tr}\,{\rm exp}\left(-\beta\hat{{\cal H}}\right)\Bigg]
\nonumber\\
\!\!\!\!&=&\!\!\!\!
-k_{\rm B}T\!\lim_{N\to\infty}\frac{1}{N}\ln\Bigg[\!\sum_{\{\sigma_k\!\}}\prod_{k=1}^N\mathrm{Tr}_k\,{\rm exp}\left(-\beta\hat{{\cal H}}_k\right)\Bigg]
\nonumber\\
\!\!\!\!&=&\!\!\!\!
- k_{\rm B}T\ln\bigg\{v(1) +\! v(-1) + \!\!\sqrt{[v(1) \!-\! v(-1)]^2 \!+\! 4v(0)^2}\bigg\},
\end{eqnarray}
where
\begin{eqnarray}
\label{eq:w}
\hspace{-2mm}
v(x) \!\!\!\!&=&\!\!\!\! \frac{1 + z^6}{2}\,{\rm exp}\left(\frac{\beta H_Ix}{2}\right)
\nonumber\\
\hspace{-2mm}
\!\!\!&&\!\!\!\!+
\Bigg\{(z\!+\!z^5)\left[{\rm exp}\left(2\beta t\right) \!+\! 2{\rm exp}\left(-\beta t\right)\right]\cosh\left[\frac{\beta (Jx\!+\! H_e)}{2}\right]
\nonumber\\
\hspace{-2mm}
\!\!\!&&\!\!\!\!+
(z^2\!+\!z^4)\left[{\rm exp}\left(-2\beta t\right) \!+\! 2{\rm exp}\left(\beta t\right)\right]\cosh\left[\beta (Jx\!+\! H_e)\right]
\nonumber\\
\hspace{-2mm}
\!\!\!&&\!\!\!\!+
\frac{z^2\!+\!z^4}{2}\Big[4{\rm exp}\left(-2\beta t\right) \!+\! 4{\rm exp}\left(\beta t\right) \!+\! {\rm exp}\left(4\beta t\right)\Big]
\nonumber\\
\hspace{-2mm}
\!\!\!&&\!\!\!\!+
z^3\cosh\left[\frac{3\beta (Jx\!+\! H_e)}{2}\right] \!+\! 5z^3\cosh\left[\frac{\beta (Jx\!+\! H_e)}{2}\right]
\nonumber\\
\hspace{-2mm}
\!\!\!&&\!\!\!\!+
4z^3\cosh\left(3\beta t\right)\cosh\left[\frac{\beta (Jx\!+\! H_e)}{2}\right]\Bigg\}\,{\rm exp}\left(\frac{\beta H_Ix}{2}\right).
\end{eqnarray}
For more computational details see Ref.~\cite{Cis14}, which deals with similar hybrid spin-electron model having the linear chain structure. The summation $\sum_{\{\sigma_k \}}$ emerging in Eq.~(\ref{eq:Omega}) includes all possible states of the localized Ising spins, the symbol $\mathrm{Tr}$ labels for a trace over the degrees of freedom of all mobile electrons in the model and $\mathrm{Tr}_k$ stands for the partial trace running over degrees of freedom of electrons from the $k$th triangular plaquette. The symbol~$\beta=1/(k_{\rm B}T)$ represents the inverse temperature ($k_{\rm B}$ is Boltzmann constant) and the parameter $z = \exp(\beta\mu)$ appearing  in Eq.~(\ref{eq:w}) denotes the electron fugacity.

\section{Ground state}
\label{sec:3}

In this section we will discuss a diversity of the ground state in dependence on interaction parameters of the model and the applied magnetic field. Due to fundamental differences between magnetic behavior of the system with distinct nature (sign) of the spin-electron coupling $J$, the analysis will be divided into two parts; the case of the ferromagnetic ($J>0$) and antiferromagnetic ($J<0$) exchange interaction. To reduce the number of free parameters, equal 'effective' magnetic fields acting on the Ising spins and electrons $H_I=H_e=H$ will be assumed in both the cases.

\subsection{The ferromagnetic exchange interaction $J>0$}
\label{subsec:31}

\begin{figure*}[t!]
\centering
\vspace{0.25cm}
\includegraphics[width = 1.0\textwidth]{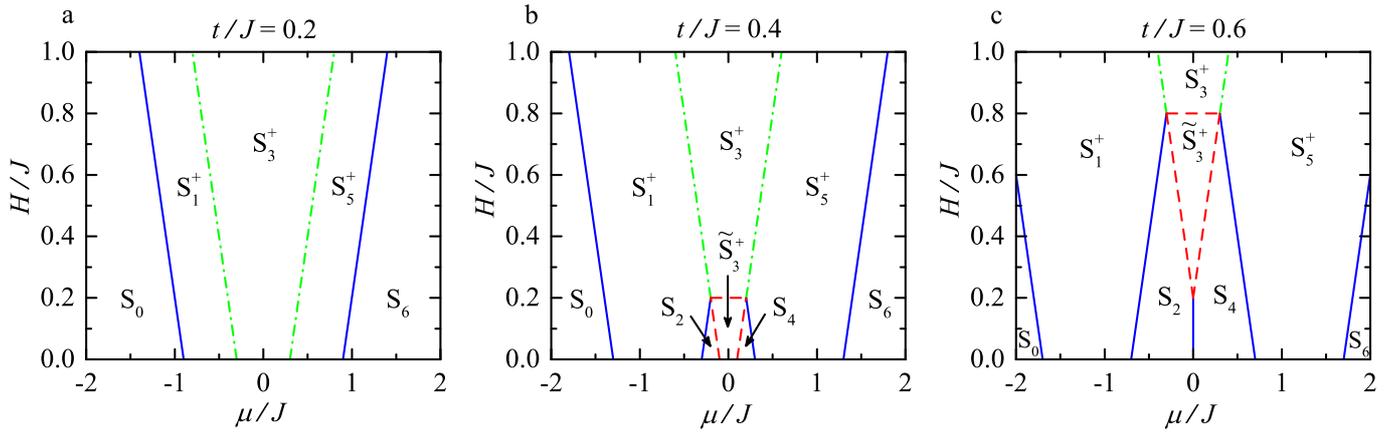}
\vspace{-0.5cm}
\caption{\small (Color online) Ground-state phase diagrams of the model with the ferromagnetic spin-electron interaction $J>0$ constructed in the $\mu-H$ plane for three representative values of the hopping term. The blue solid, red dashed and green dot-dashed lines denote first-order phase transitions with the macroscopic degeneracies ${\cal W} = 2^N$, $3^N$ and $4^N$, respectively.}
\label{fig:2}
\end{figure*}

Typical ground-state phase diagrams of the model with the ferromagnetic spin-electron interaction $J>0$ in the $\mu-H$ plane are illustrated in Fig.~\ref{fig:2}. Obviously, the displayed phase diagrams are symmetric with respect to the line $\mu = 0$ and contain eight different ground-state phases, which can be characterized by the following eigenvectors and ground-state energies:
\begin{eqnarray}
\hspace{-6mm}
\label{eq:S_0,2,4,6}
|{\rm S}_{a}\rangle \!\!\!\!&=&\!\!\!\! \prod_{k=1}^N|{\rm S}_{a}\rangle_k  = \prod_{k=1}^N\left\{  \!\!\begin{tabular}{l}
                                                            $|(\downarrow)\!\uparrow\rangle_{\sigma_k}$
                                                            $\!\otimes$
                                                            $|n_{k}=a,S_{k}^{z}=0\rangle$,\, $H = 0$
                                                            \\[1mm]
                                                            $|\!\uparrow\rangle_{\sigma_k}$ $\!\otimes$
                                                            $|n_{k}=a,S_{k}^{z}=0\rangle$\hspace{3mm}  ,\, $H > 0$
                                                     \end{tabular}
                                        \right.\!\!\!\!,
                                        \nonumber\\
                                        \hspace{-6mm}
E_{a} \!\!\!\!&=&\!\!\!\! -\frac{N}{2}\left[ H + 2a\mu + a(6-a)t\right],\,\,\, a=\{0,2,4,6\};
\\[1mm]
\hspace{-6mm}
\label{eq:S_1,5+}
|{\rm S}_{b}^{+}\rangle \!\!\!\!&=&\!\!\!\! \prod_{k=1}^N|{\rm S}_{b}^{+}\rangle_k =
\prod_{k=1}^N\left\{      \!\!\begin{tabular}{l}
                                                            $|\!\downarrow\rangle_{\sigma_k}$ $\!\otimes$
                                                            $|n_{k}=b,S_{k}^{z}=-\frac{1}{2}\rangle$ ,\,
                                                            $H = 0$
                                                            \\[1mm]
                                                            $|\!\uparrow\rangle_{\sigma_k}$ $\!\otimes$
                                                            $|n_{k}=b,S_{k}^{z}=\frac{1}{2}\rangle$\hspace{3mm}  \hspace{-1mm}, $H \geq 0$
                                                     \end{tabular}
                                        \right.\!\!\!\!,
                                        \nonumber\\
                                        \hspace{-6mm}
E_{b}^{+} \!\!\!\!&=&\!\!\!\! -\frac{N}{2}\left( J + 2H +2b\mu + 4t \right),\,\,\, b=\{1,5\};
\\[1mm]
\hspace{-6mm}
\label{eq:S_3+}
|{\rm S}_{3}^{+}\rangle \!\!\!\!&=&\!\!\!\! \prod_{k=1}^N|{\rm S}_{3}^{+}\rangle_k =
\prod_{k=1}^N\left\{ \!\!\begin{tabular}{l}
                                                            $|\!\downarrow\rangle_{\sigma_k}$ $\!\otimes$
                                                            $|n_{k}=3,S_{k}^{z}=-\frac{3}{2}\rangle$ ,\,
                                                            $H = 0$
                                                            \\[1mm]
                                                            $|\!\uparrow\rangle_{\sigma_k}$ $\!\otimes$
                                                            $|n_{k}=3,S_{k}^{z}=\frac{3}{2}\rangle$\hspace{7mm}  \hspace{-5mm}, $H \geq 0$
                                                     \end{tabular}
                                               \right.\!\!\!\!,
                                                \nonumber\\
                                                \hspace{-6mm}
E_{3}^{+} \!\!\!\!&=&\!\!\!\! -\frac{N}{2}\left(3J + 4H + 6\mu\right);
\\[1mm]
\hspace{-6mm}
\label{eq:S_3chiral+}
|{\rm \widetilde{S}}_{3}^{+}\rangle \!\!\!\!&=&\!\!\!\! \prod_{k=1}^N|{\rm \widetilde{S}}_{3}^{+}\rangle_k =
\prod_{k=1}^N\left\{ \!\!\begin{tabular}{l}
                                                            $|\!\downarrow\rangle_{\sigma_k}$ $\!\otimes$
                                                            $|n_{k}=3,S_{k}^{z}=-\frac{1}{2}\rangle_{L, R}$,\,
                                                            $H = 0$
                                                            \\[1mm]
                                                            $|\!\uparrow\rangle_{\sigma_k}$ $\!\otimes$
                                                            $|n_{k}=3,S_{k}^{z}=\frac{1}{2}\rangle_{ L, R}$\hspace{2mm}, $H \geq 0$
                                                     \end{tabular}
                                               \right.\!\!\!\!,
                                                \nonumber\\
                                                \hspace{-6mm}
\widetilde{E}_{3}^{+} \!\!\!\!&=&\!\!\!\! -\frac{N}{2}\left(J + 2H + 6\mu + 6t\right).
\end{eqnarray}
In above, the product~$\prod_{k=1}^{N}$ runs over all elementary unit cells, the state vector $|\!\uparrow\rangle_{\sigma_k}$ ($|\!\downarrow\rangle_{\sigma_k}$) determines up (down) state of the Ising spin at the $k$th nodal lattice site, while $|n_{k}, S^{z}_{k}\rangle$, $|n_{k}, S^{z}_{k}\rangle_{L,R}$ refer to the non-chiral and chiral eigenstates of $n_{k}$ mobile electrons in the $k$th triangular plaquette with the total spin $S^{z}_{k}$, respectively. Analytical expressions of the state vectors $|n_{k}, S^{z}_{k}\rangle$, $|n_{k}, S^{z}_{k}\rangle_{L,R}$ are too cumbersome for some electron concentrations, therefore we omit them in this work. Finally, the subscripts $a$, $b$, $3$ specify a total number of electrons per triangular plaquette in individual phases, and the superscript 'plus' in the phases~(\ref{eq:S_1,5+})--(\ref{eq:S_3chiral+}) points out to the same signs of the total spin of electron plaquette and the state of the Ising spin that form $k$th elementary unit, i.e. it is a result of the product ${\rm sgn}(S_k^z)\cdot{\rm sgn}(\sigma_k^z)$.

As can be understood from Eqs.~(\ref{eq:S_0,2,4,6})--(\ref{eq:S_3chiral+}), the arrangement of the Ising spins depends on a presence of the magnetic field and the total spin of triangular plaquettes. To be specific, the first group of phases S$_{a}$, within which triangular plaquettes have zero total spin, because they are either empty, fully-filled or partly occupied by two or four electrons in a quantum superposition of several non-magnetic and intrinsic antiferromagnetic states, exhibits a kinetically-driven frustration of the Ising spins at the zero magnetic field. As a result, the residual entropy per elementary unit ${\cal S} = k_{\rm B}\!\ln 2$ can be observed for $H=0$, which indicates the zero-field macroscopic degeneracy ${\cal W} = 2^N$. An arbitrary non-zero field cancels this degeneracy, because it forces all the Ising spins to align into its direction.
On the other hand, if triangular plaquettes are occupied by odd number of electrons, as it is in the phases S$_{b}^{+}$ and S$_{3}^{+}$, all the Ising spins in the model may occupy either the down or up states with the same probability if $H=0$. Mobile electrons in triangular plaquettes also choose between two possible quantum states with the total spins $S_{k}^{z}= -1/2$ and $1/2$ (in S$_{b}^{+}$) or between two classical ferromagnetic states corresponding to the total spins $S_{k}^{z}= -3/2$ and $3/2$  (in S$_{3}^{+}$) in order to preserve the spontaneous ferromagnetic order with the nearest Ising neighbors. Any non-zero magnetic field lifts this two-fold degeneracy of the phases S$_{b}^{+}$, S$_{3}^{+}$. Last but not least, there can also be found a peculiar field-independent macroscopic degeneracy if the phase ${\rm \widetilde{S}}_{3}^{+}$ constitutes the ground state. It is valuable to point out that ${\rm \widetilde{S}^{+}}_{3}$ is the only one ground state from the group of phases~(\ref{eq:S_0,2,4,6})--(\ref{eq:S_3chiral+}), in which the entropy per unit cell remains non-zero in the whole parameter space. The observed value ${\cal S} = k_{\rm B}\!\ln 2$ is a result of two possible chiral degrees of freedom ($L$eft- and $R$ight-hand side) of mobile electrons in each triangular plaquette. We note that the phase ${\rm \widetilde{S}}_{3}^{+}$ and the phases S$_{2}$, S$_{4}$ can be identified in the ground state just for the hopping terms~$t/J>1/3$ (see Fig.~\ref{fig:2}).

In general, the high macroscopic degeneracy can also be found at points of discontinuous (first-order) phase transitions, where different ground states are in thermodynamic equilibrium. Positions of these points can be identified by comparing the ground-state energies of individual phases. Namely, macroscopic degeneracies of three different sizes ${\cal W} = 2^N$, $3^N$ and $4^N$ may be observed along individual phase boundaries. The first degeneracy can be found along the phase transitions S$_2$--S$_4$ and S$_p^{+}$--S$_{p\pm1}$ ($p=\{1, 5\}$), while the second one
corresponds to the boundaries that separate the chiral phase ${\rm \widetilde{S}^{+}}_{3}$ from the neighboring non-chiral ones S$_2$, S$_4$ and S$_3^{+}$. The last (highest) degeneracy can be found at two special phase transitions S$_q^{+}$--S$_{q+2}^{+}$ ($q=\{1,3\}$), which are marked as green dot-dashed lines in Fig.~\ref{fig:2}. At each point of these boundaries, spin-electron configurations of elementary unit cells corresponding to the neighboring phases S$_q^{+}$, S$_{q+2}^{+}$ are in thermodynamic equilibrium with another (novel) configuration ${\rm \widetilde{S}}_{(q+1)k}^{+}$ given by the following eigenvector and ground-state energy
\begin{eqnarray}
\label{eq:S_2,4chiral+}
|{\rm \widetilde{S}}_{q+1}^{+}\rangle_{k} \!\!\!\!&=&\!\!\!\! \left\{ \begin{tabular}{l}
                                                            $|\!\downarrow\rangle_{\sigma_k}$ $\!\otimes$
                                                            $|n_{k}=q+1,S_{k}^{z}=-1\rangle_{L, R}$,\,\,\,
                                                            $H = 0$
                                                            \\[1mm]  $|\!\uparrow\rangle_{\sigma_k}$ $\!\otimes$
                                                            $|n_{k}=q+1,S_{k}^{z}=1\rangle_{ L, R}$\hspace{2mm},\hspace{2mm}\,$H \geq 0$
                                                     \end{tabular}
                                               \right.\!\!\!\!,
                                                \nonumber\\
\widetilde{E}_{(q+1)k}^{+} \!\!\!\!&=&\!\!\!\! -J -\frac{3H}{2} - (q+1)\mu  - t.
\end{eqnarray}
Similarly as the phase ${\rm \widetilde{S}}_{3}^{+}$, the configurations~(\ref{eq:S_2,4chiral+}) are macroscopically degenerate due to the chiral degrees of freedom of mobile electrons. Their existence along the phase boundaries  S$_q^{+}$--S$_{q+2}^{+}$ causes the double increase in the macroscopic degeneracy compared to the boundaries, where two non-chiral ground states are in thermodynamic equilibrium.

\subsection{The antiferromagnetic exchange interaction $J<0$}
\label{subsec:32}

In agreement with general expectations, more complex ground-state behavior can be observed for the model with the antiferromagnetic spin-electron interaction $J<0$. In fact, a~mutual competition between the parameters $t>0$, $J<0$, $\mu$ and $H\geq0$ generates three novel non-chiral and one novel chiral magnetic phases at the zero temperature apart from those found for the ferromagnetic version of model. These phases are unambiguously characterized by the following eigenvectors and ground-state energies:
\begin{eqnarray}
\hspace{-6mm}
\label{eq:S_1,5-}
|{\rm S}_{b}^{-}\rangle \!\!\!\!&=&\!\!\!\! \prod_{k=1}^N|{\rm S}_{b}^{-}\rangle_k =
\prod_{k=1}^N\left\{      \!\!\begin{tabular}{l}
                                                            $|\!\uparrow\rangle_{\sigma_k}$ $\!\otimes$
                                                            $|n_{k}=b,S_{k}^{z}=-\frac{1}{2}\rangle$ \\[1mm]
                                                            $|\!\downarrow\rangle_{\sigma_k}$ $\!\otimes$
                                                            $|n_{k}=b,S_{k}^{z}=\frac{1}{2}\rangle$\hspace{3mm}
                                \end{tabular}
                                        \right.\hspace{-4mm},\,H \geq 0,
                                        \nonumber\\
                                        \hspace{-6mm}
E_{b}^{-} \!\!\!\!&=&\!\!\!\! \frac{N}{2}\left( J -2b\mu - 4t \right),\,\,\, b=\{1,5\};
\\[1mm]
\hspace{-6mm}
\label{eq:S_3-}
|{\rm S}_{3}^{-}\rangle \!\!\!\!&=&\!\!\!\! \prod_{k=1}^N|{\rm S}_{3}^{-}\rangle_k =
\prod_{k=1}^N\left\{ \!\!\begin{tabular}{l}
                                                            $|\!\uparrow\rangle_{\sigma_k}$ $\!\otimes$
                                                            $|n_{k}=3,S_{k}^{z}=-\frac{3}{2}\rangle$ ,\,
                                                            $H = 0$
                                                            \\[1mm]
                                                            $|\!\downarrow\rangle_{\sigma_k}$ $\!\otimes$
                                                            $|n_{k}=3,S_{k}^{z}=\frac{3}{2}\rangle$\hspace{7mm}  \hspace{-5mm}, $H \geq 0$
                                                     \end{tabular}
                                               \right.\!\!\!\!,
                                                \nonumber\\
                                                \hspace{-6mm}
E_{3}^{-} \!\!\!\!&=&\!\!\!\! \frac{N}{2}\left(3J - 2H - 6\mu\right);
\\[1mm]
\hspace{-6mm}
\label{eq:S_3chiral-}
|{\rm \widetilde{S}}_{3}^{-}\rangle \!\!\!\!&=&\!\!\!\! \prod_{k=1}^N|{\rm \widetilde{S}}_{3}^{-}\rangle_k =
\prod_{k=1}^N\left\{ \!\!\begin{tabular}{l}
                                                            $|\!\uparrow\rangle_{\sigma_k}$ $\!\otimes$
                                                            $|n_{k}=3,S_{k}^{z}=-\frac{1}{2}\rangle_{L, R}$
                                                            \\[1mm]
                                                            $|\!\downarrow\rangle_{\sigma_k}$ $\!\otimes$
                                                            $|n_{k}=3,S_{k}^{z}=\frac{1}{2}\rangle_{ L, R}$
                                                     \end{tabular}
                                               \right.\hspace{-3mm},\,H \geq 0,
                                                \nonumber\\
                                                \hspace{-6mm}
\widetilde{E}_{3}^{-} \!\!\!\!&=&\!\!\!\! \frac{N}{2}\left(J - 6\mu - 6t\right).
\end{eqnarray}
In above, the superscript 'minus' is again a result of the product ${\rm sgn}(S_k^z)\cdot{\rm sgn}(\sigma_k^z)$.

\begin{figure*}[t!]
\centering
\vspace{0.25cm}
\includegraphics[width = 1.0\textwidth]{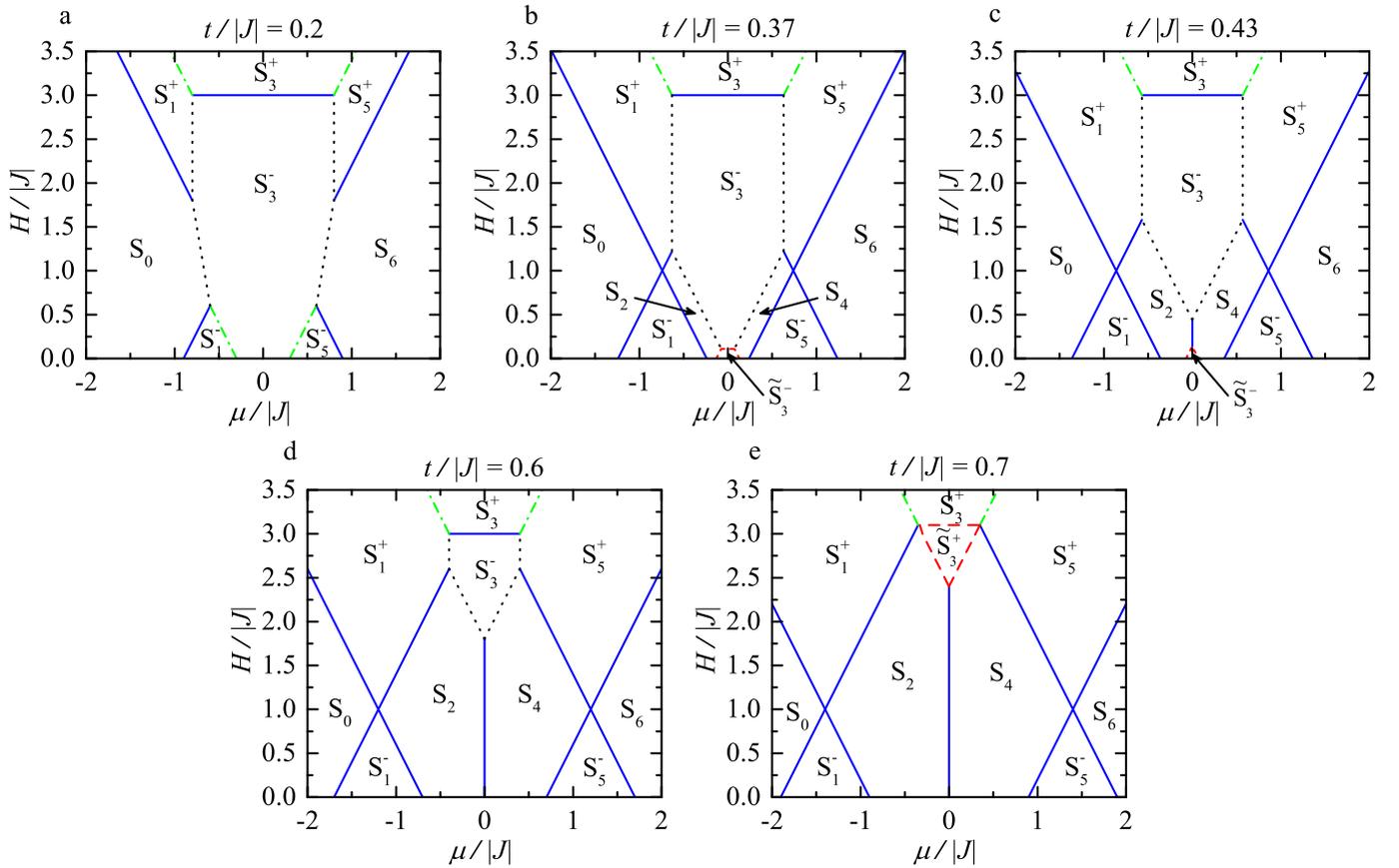}
\vspace{-0.5cm}
\caption{\small (Color online) Ground-state phase diagrams of the model with the antiferromagnetic spin-electron interaction $J<0$ constructed in the $\mu-H$ plane for five representative values of the hopping parameter. The black dotted lines denote first-order phase transitions, where the system is two-fold degenerate, while blue solid, red dashed and green dot-dashed lines label first-order phase transitions with the macroscopic degeneracies ${\cal W} = 2^N$, $3^N$ and $4^N$, respectively.}
\label{fig:3}
\end{figure*}

The ground-state diagram in the $\mu-H$ plane can have several topologies in dependence on the hopping parameter (see~Fig.~\ref{fig:3}). As can be found from Fig.~\ref{fig:3}a, the zero-temperature $\mu-H$ plane includes solely non-chiral phases, namely S$_0$, S$_1^{+}$, S$_3^{+}$, S$_5^{+}$, S$_6$, which have already been observed in the ferromagnetic counterpart of the model, and three new S$_b^{-}$, S$_3^{-}$ if $t/|J|<1/3$. It is clear from Eqs.~(\ref{eq:S_1,5-}), (\ref{eq:S_3-}) that the latter group of phases are characterized by  opposite signs of the total spins of electron plaquettes and the Ising spin states. Moreover, the phases S$_b^{-}$, which appear in the low-field region $H/|J|<1$, remain two-fold degenerate in the whole parameter space. By contrast, the two-fold degeneracy of the phase S$_3^{-}$ is completely canceled as soon as the magnetic field is switched on. If the reverse condition $t/|J|>1/3$ holds, two other non-chiral phases S$_2$, S$_4$ and two chiral phases ${\rm \widetilde{S}}_3^{-}$, ${\rm \widetilde{S}}_3^{+}$ may be observed in the $\mu-H$ plane. The effect of the hopping parameter on the evolution of all four phases can be understood from Figs.~\ref{fig:3}b--e. Obviously, stability regions of the non-chiral phases S$_2$, S$_4$, which can also be found for the ferromagnetic spin-electron interaction $J>0$, are gradually growing at the expense of the phase S$_3^{-}$ when $t$ becomes stronger. The novel chiral phase ${\rm \widetilde{S}}_3^{-}$ emerges merely between the phases S$_2$, S$_4$ in the weak-field region for $1/3<t/|J|<1/2$. If $1/3<t/|J|<2/5$, the horizontal field-induced phase transition ${\rm \widetilde{S}}_3^{-}$--S$_3^{-}$ is gradually shorten and shifted towards higher (but still very weak) magnetic fields with the increasing kinetic term until it completely disappears at $t/|J|=2/5$. If the value of the parameter $t$ further increases, a new horizontal phase boundary S$_2$--S$_4$ emerges along the line $\mu=0$. As a consequence, the phase ${\rm \widetilde{S}}_3^{-}$ completely disappears from the ground-state phase diagram after exceeding the value $t/|J| = 1/2$. Further strengthening of the hopping parameter, which prefers an antiparallel orientation of mobile electrons and the jumping of these particles between individual plaquette sites, leads to the suppression of the interaction $J<0$. As a result, the non-chiral phase S$_3^{-}$, in which electron plaquettes are fully polarized into the magnetic-field direction and the Ising spins are in opposite orientation with respect to the nearest-neighboring electrons, transforms to the chiral one~${\rm \widetilde{S}}_3^{+}$ at $t/|J| = 2/3$. In fact, the phase ${\rm \widetilde{S}}_3^{+}$ is energetically more favorable than S$_3^{-}$ for any values $t/|J|>2/3$ (see Fig.~\ref{fig:3}e).

Regarding to the macroscopic degeneracy of the system, this phenomenon can be identified in the chiral phases ${\rm \widetilde{S}}_3^{+}$, ${\rm \widetilde{S}}_3^{-}$, the non-chiral phases S$_a$ if the magnetic field is zero, and also at some first-order phase transitions.
It is clear from Eqs.~(\ref{eq:S_3chiral+}), (\ref{eq:S_3chiral-}), that the common feature of the first two phases ${\rm \widetilde{S}}_3^{+}$, ${\rm \widetilde{S}}_3^{-}$ is the field-independent two-fold degeneracy of each electron plaquette due to two possible values of the scalar chirality, which leads to the macroscopic degeneracy proportional to $2^N$. Moreover, the phase ${\rm \widetilde{S}}_3^{-}$ is two-fold degenerate in the whole parameter space with respect to two possible spin-electron configurations with opposite signs of $S_k^z$ and $\sigma_k^z$.  As a result, the field-independent macroscopic degeneracy of the phase ${\rm \widetilde{S}}_3^{-}$ is a little bit higher compared to the one of ${\rm \widetilde{S}}_3^{+}$;  one can find ${\cal W} = 2^{N+1}$ in ${\rm \widetilde{S}}_3^{-}$ and  ${\cal W} = 2^{N}$ in ${\rm \widetilde{S}}_3^{+}$. According to the ground-state analysis, the macroscopic degeneracy ${\cal W} = 2^{N+1}$ appears in the zero-temperature $\mu-H$ plane for the hopping parameters $1/3<t/|J|<1/2$, while the macroscopic degeneracy ${\cal W} = 2^{N}$ can be observed for $t/|J|>2/3$. It is worthy to recall that the macroscopic degeneracy of the size ${\cal W} = 2^{N}$ can also be detected in the zero-field region if the phases S$_a$ constitute the ground state. However, this degeneracy is completely canceled by the applied magnetic field, because it comes from the kinetically-driven frustration of the Ising spins [see Eq.~(\ref{eq:S_0,2,4,6})].

Last but not least, the model with the spin-electron interaction $J<0$ is  macroscopically degenerate also at most first-order phase transitions. One can conclude from Fig.~\ref{fig:3} that degeneracy degrees are equal to those observed for the ferromagnetic counterpart of the model: the set of phase transitions S$_p^{\alpha}$--S$_{p\pm1}$ ($\alpha =\{\pm\}$), S$_2$--S$_4$, S$_3^{-}$--S$_3^{+}$ (blue solid  lines) is associated with the degeneracy ${\cal W} = 2^N$, the boundaries separating the chiral phases ${\rm \widetilde{S}^{\alpha}}_{3}$ from the non-chiral ones S$_{3}^{\alpha}$, S$_{2}$, S$_{4}$ (red dashed lines) have the degeneracy ${\cal W} = 3^N$, and finally, the phase transitions S$_q^{\alpha}$--S$_{q+2}^{\alpha}$ (green dot-dashed lines) exhibit the degeneracy ${\cal W} = 4^N$. Note that the macroscopic degeneracy observed at boundaries between the phases S$_q^{\alpha}$, S$_{q+2}^{\alpha}$ is caused by a mutual coexistence of the elementary-unit configurations peculiar to neighboring phases with the chiral ones, namely ${\rm \widetilde{S}}_{(q+1)k}^{+}$ given by Eq.~(\ref{eq:S_2,4chiral+}) (in S$_q^{+}$--S$_{q+2}^{+}$) or ${\rm \widetilde{S}}_{(q+1)k}^{-}$ (in \mbox{S$_q^{+}$--S$_{q+2}^{+}$}). The latter chiral configurations of the elementary unit cell are given by
\begin{eqnarray}
\label{eq:S_2,4chiral-}
|{\rm \widetilde{S}}_{q+1}^{-}\rangle_{k} \!\!\!\!&=&\!\!\!\! \left\{ \begin{tabular}{l}
                                                            $|\!\uparrow\rangle_{\sigma_k}$ $\!\otimes$
                                                            $|n_{k}=q+1,S_{k}^{z}=-1\rangle_{L, R}$,\,\,\,
                                                            $H = 0$
                                                            \\[1mm]  $|\!\downarrow\rangle_{\sigma_k}$ $\!\otimes$
                                                            $|n_{k}=q+1,S_{k}^{z}=1\rangle_{ L, R}$\hspace{2mm},\hspace{2mm}\,$H \geq 0$
                                                     \end{tabular}
                                               \right.\!\!\!\!,
                                                \nonumber\\
\widetilde{E}_{(q+1)k}^{-} \!\!\!\!&=&\!\!\!\! J -\frac{H}{2} - (q+1)\mu  - t.
\end{eqnarray}
For completeness, it is also necessary to describe a degeneracy of not yet mentioned phase transitions S$_3^{-}$--S$_a$ and S$_3^{-}$--S$_b^{+}$, which are marked as dotted lines in Fig.~\ref{fig:3}. The investigated model is two-fold degenerate at each point of these transitions, because only one localized Ising spin in the model is free to flip, while spin states of the other ones are uniquely determined by mutual interplay between the parameters $J<0$,~$H>0$. It should be noted that the two-fold degeneracy of the first-order phase transitions has not been observed in the ground state of the model with the ferromagnetic spin-electron coupling~$J>0$.

\section{Low-temperature thermodynamics}
\label{sec:4}

\begin{figure*}[t!]
\centering
\vspace{0.25cm}
\includegraphics[width = 1.0\textwidth]{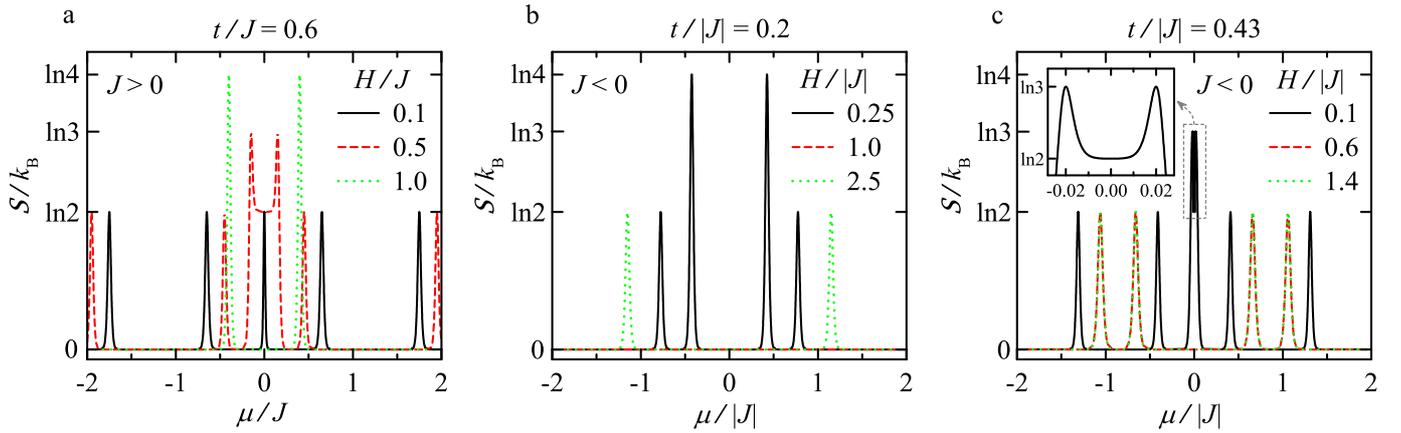}
\vspace{-0.5cm}
\caption{\small (Color online) The entropy per elementary unit cell as a function of the chemical potential for the model with the ferromagnetic (figure~a) as well as the antiferromagnetic  (figures~b, c) spin-electron interaction $J$ and several chosen combinations of the parameters $t$, $H$. The~temperature is fixed to the value $k_{\rm B}T/J = k_{\rm B}T/|J| = 0.01$.}
\label{fig:4}
\end{figure*}

The macroscopic ground-state degeneracy may manifest itself in the low-temperature behavior of basic thermodynamic quantities such as entropy and specific heat. The entropy per elementary unit cell in relevant ground-state phase transitions can be directly determined from the degeneracy of the ground-state manifold according to the formula ${\cal S} = k_{\rm B}\!\lim\limits_{N\to\infty}N^{-1}\ln{\cal W}$~\cite{Der04}. Bearing in mind the results presented in Sec.~\ref{sec:3}, the residual entropy per unit cell may take three finite values at the boundaries between different ground states: ${\cal S} = k_{\rm B}\!\ln 2\approx0.693k_{\rm B}$ at S$_2$--S$_4$, S$_3^{-}$--S$_3^{+}$, S$_p^{\alpha}$--S$_{p\pm1}$, ${\cal S} = k_{\rm B}\!\ln 3\approx1.099k_{\rm B}$ at ${\rm \widetilde{S}^{\alpha}}_{3}$--S$_{3}^{\alpha}$, \mbox{${\rm \widetilde{S}^{\alpha}}_{3}$--S$_{2}$,} ${\rm \widetilde{S}^{\alpha}}_{3}$--S$_{4}$, and ${\cal S} = k_{\rm B}\!\ln 4\approx1.386k_{\rm B}$ at S$_q^{\alpha}$--S$_{q+2}^{\alpha}$ (recall that $p=\{1, 5\}$, $q=\{1, 3\}$ and $\alpha =\{\pm\}$). An exception is the group of phase transitions S$_3^{-}$--S$_a$, S$_3^{-}$--S$_b^{+}$ observed in the ground state of the model with the spin-electron coupling $J<0$ (see Fig.~\ref{fig:3}). The entropy of the system takes the finite value $\ln 2$ at each point of these phase transitions due to its two-fold degeneracy caused by one free Ising spin. However, this contribution vanishes in the thermodynamic limit $N\to\infty$. Consequently, the entropy normalized per unit cell takes the value ${\cal S} = 0$, which clearly implies that the model is not macroscopically degenerate at the phase transitions S$_3^{-}$--S$_a$, S$_3^{-}$--S$_b^{+}$. The afore-mentioned statements can be independently checked by low-temperature variations of the entropy depicted in Fig.~\ref{fig:4}. We note that the plotted curves have been obtained from the thermodynamic relation ${\cal S} = -\left(\partial \Omega/\partial T\right)_{H, J, \mu}$~\cite{Hua63}. As one sees in Fig.~\ref{fig:4}, the low-temperature entropy as a function of the chemical potential exhibits a series of narrow peaks at the values corresponding to the relevant ground-state phase transitions, excepting to those that are associated with the phase boundaries S$_3^{-}$--S$_a$, S$_3^{-}$--S$_b^{+}$. The magnitudes of observed peaks are in accordance with the previously reported values of the residual entropy normalized per unit cell. Moreover, two plateaux at ${\cal S} = k_{\rm B}\!\ln 2$ can be identified for the combinations of model parameters $t/J=0.6$, $H/J=0.5$ (if $J>0$) and $t/|J|=0.43$, $H/|J|=0.1$ (if $J<0$) between peaks of the heights ${\cal S} = k_{\rm B}\!\ln 3$ (see Figs.~\ref{fig:4}a, c). 
Referring to the phase diagrams plotted in Figs.~\ref{fig:2}c and~\ref{fig:3}c one can conclude that these plateaux correspond to the macroscopically degenerate chiral phases ${\rm \widetilde{S}^{+}}_{3}$ and ${\rm \widetilde{S}^{-}}_{3}$, respectively.

The solution for the grand-canonical potential~(\ref{eq:Omega}) allows one to exactly examine the degeneracy of the model also in terms of low-temperature variations of the specific heat. This physical quantity can be obtained by using the relation \mbox{${\cal C} = -T(\partial^{2} \Omega/\partial T^{2})_{H, J, \mu}$~\cite{Hua63}.}

\begin{figure*}[t!]
\centering
\vspace{0.25cm}
\includegraphics[width = 1.0\textwidth]{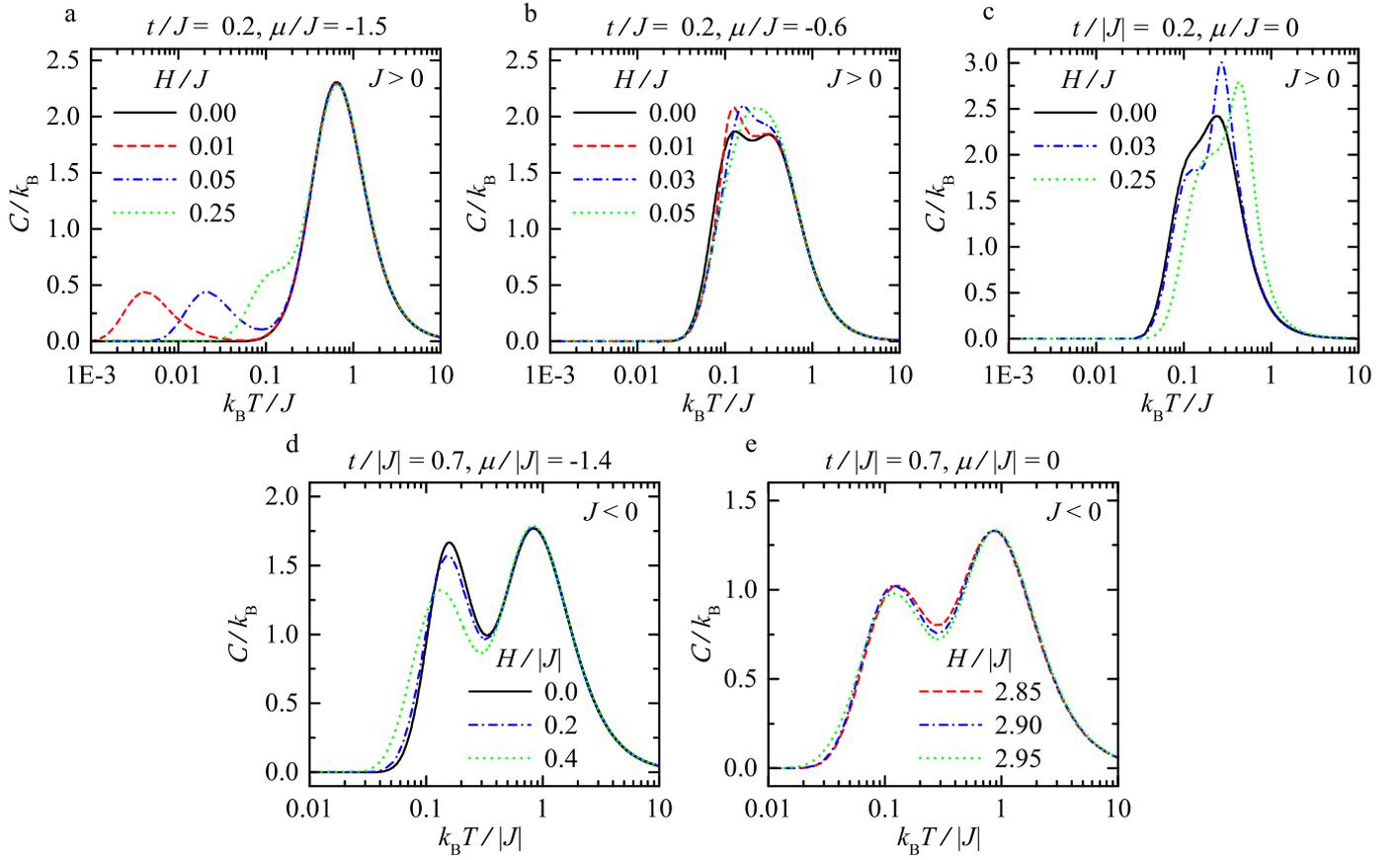}
\vspace{-0.5cm}
\caption{\small (Color online) The field evolution of the specific heat versus temperature if the zero-field ground state is constituted by the phase S$_0$ (figure~a), the phases S$_1^{+}$ (figure~b), S$_3^{+}$ (figure~c), S$_1^{-}$ (figure~d), and if the chiral phase ${\rm \widetilde{S}}_{3}^{+}$ is stable the ground state~(figure~e).}
\label{fig:5}
\end{figure*}

Figure~\ref{fig:5} illustrates the effect of the external magnetic field on temperature dependencies of the specific heat when the ground state is constituted by the phases of different degeneracies. As one sees from Fig.~\ref{fig:5}a, if the zero-field ground state is macroscopically degenerate due to frustration of the Ising spins, the standard single-peak structure of the specific observed for $H=0$ changes to the double-peak one at relatively weak fields. The formation of an additional Schottky-type maximum in the low-temperature part of ${\cal C}(T)$ curve can be attributed to Zeeman splitting of energy levels of the frustrated Ising spins. In accordance with this statement, the field-induced low-temperature maximum gradually shifts towards higher temperatures with the increasing $H$ until it merges with the high-temperature maximum. For two-fold degeneracy that disappears at $H>0$, the zero-field ${\cal C}(T)$ curve may have two-peak or irregular single-peak structure depending on whether total spins of electron plaquettes are $S_k^z = \pm1/2$ or $S_k^z = \pm3/2$, respectively. In the former case, the increasing external field causes a gradual rise of the low-temperature peak and its shifting towards higher temperatures until it completely coalesces with the high-temperature maximum (see Fig.~\ref{fig:5}b). The field evolution of the irregular single-peak structure corresponding to $S_k^z = \pm3/2$ is different. It firstly changes to the two-peak structure at very weak magnetic fields. However, a few significant low-temperature peak very quickly merge with the high-temperature maximum with further increase of $H$ (see Fig.~\ref{fig:5}c). Finally, if two-fold and macroscopic degeneracies of the ground state are maintained at finite magnetic fields, as it is in the phases S$_{b}^{-}$, ${\rm \widetilde{S}}_3^{\alpha}$, two pronounced Shottky-type maxima can be identified in  temperature dependencies of the zero-field specific heat (see Figs.~\ref{fig:5}d, e). In accordance with general expectations, the observed double-peak structure does not significantly change with the increasing magnetic field if one considers parameter combination sufficiently far from phase boundaries.

\begin{figure*}[t!]
\centering
\vspace{0.25cm}
\includegraphics[width = 1.0\textwidth]{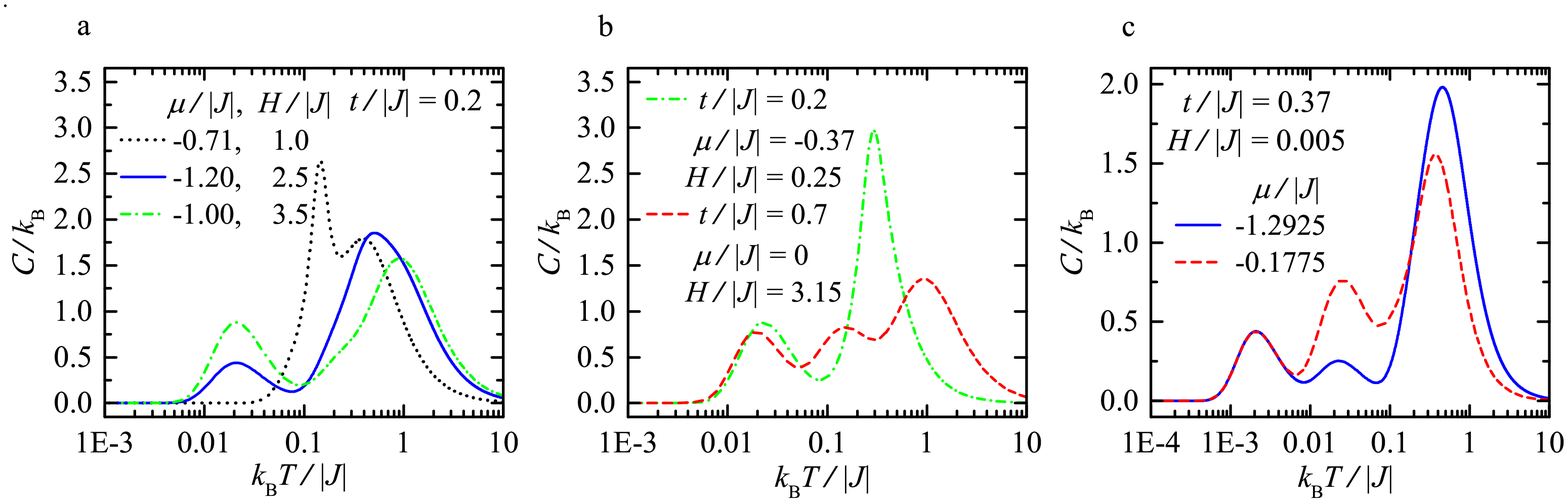}
\vspace{-0.5cm}
\caption{\small (Color online) Typical temperature dependencies of the specific heat of the model with the antiferromagnetic spin-electron interaction $J<0$ for the combinations of model parameters from a proximity of the first-order phase transitions S$_0$--S$_3^{-}$ (black dotted curve), S$_0$--S$_1^{+}$ (blue solid curve), S$_1^{+}$--S$_3^{+}$ (green dot-dashed curve) (figure~a), S$_1^{-}$--S$_3^{-}$ (green dot-dashed curve), ${\rm \widetilde{S}}_1^{+}$--S$_3^{+}$ (red dashed curve) (figure~b), and if the phases S$_0$ (blue solid curve), S$_4$ (red dashed curve) constitute the zero-field ground state (figure~c).}
\label{fig:6}
\end{figure*}

The interesting multiple-peak dependencies of the specific heat appear also in a proximity of the first-order phase transitions. Typical temperature variations of the quantity ${\cal C}$ corresponding to these regions are illustrated in Fig.~\ref{fig:6}.
If the set of model parameters is chosen to be close enough to boundaries that separate two ground-state phases with unique spin-electron arrangement, ${\cal C}(T)$ curve exhibits one broad Schottky-type maximum in a high-temperature region and one low-temperature Schottky-type maximum regardless of the degeneration degree of the respective phase transition (see Fig.~\ref{fig:6}a). The additional low-temperature peak originates due to thermal excitations between the ground-state configuration and the low-lying excited state with particle configuration of the neighboring phase. A~qualitatively similar temperature variation of the specific heat can also be seen in a proximity of the phase transitions between ground state with the unique particle arrangement and two-fold degenerate ground state (see green dot-dashed curve in Fig.~\ref{fig:6}b).
However, if model parameters are from a neighborhood of the transition between the uniquely ordered and macroscopically degenerate chiral phases, one observes an interesting ${\cal C}(T)$ curve with three Shottky-type peaks (see red dashed curve in Fig.~\ref{fig:6}b).
The maximum at high temperatures involves thermal excitations of diverse physical origin, while the other one located at moderate temperatures originates due to a large number of low-lying excited states that arise out from the chiral phase. Recall that both these peaks have already been identified in $C(T)$ curves corresponding to the chiral ground state ${\rm \widetilde{S}}_3^+$ (see Fig.~\ref{fig:5}e). The third novel peak, which emerges at the lowest temperatures, is a result of thermally induced excitations between particle configurations of the neighboring ground states.
A remarkable triple-peak temperature dependencies of the specific heat can also be expected in a proximity of the phase transitions S$_a$--S$_b^{-}$, S$_2$--${\rm \widetilde{S}}_3^{\alpha}$, S$_4$--${\rm \widetilde{S}}_3^{\alpha}$ if the phases S$_a$ constitute the ground state and the external magnetic field is very weak, as is demonstrated in Fig.~\ref{fig:6}c. In this case, the moderate peak emerging between two others is associated with low-lying excited states that may be either two-fold or macroscopically degenerate depending on whether they arise out from S$_b^{-}$ or $\widetilde{S}_3^{\alpha}$ ground states, respectively. On the other hand, the origin of the maximum emerging at the lowest temperatures lies in the field-induced splitting of energy levels of the frustrated Ising spins, which can be observed in the phases S$_a$ at $H=0$.

\section{Concluding remarks}
\label{sec:5}

The present paper deals with the exactly solvable double-tetrahedral chain of the localized Ising spins and mobile electrons. Assuming the equal magnetic field acting on all particles, possible ground-state configurations have been particularly discussed for ferromagnetic and antiferromagnetic exchange interactions between mobile electrons and their nearest Ising neighbors. Moreover, thermodynamic quantities such as entropy and specific heat have been calculated by using the exact analytical solution for the grand-canonical potential in order to gain a deeper insight into the degeneracy of individual ground-state phases and phase transitions.

It has been shown that the zero-temperature phase diagram of the model may involve several non-degenerate, two-fold degenerate and macroscopically degenerate chiral phases as a result of mutual interplay between model parameters and the external magnetic field. It has been also demonstrated that the macroscopic degeneracy arising from chiral degrees of freedom of mobile electrons is not canceled by the external magnetic field in contrast to the macroscopic degeneracy appearing due to a frustration of the localized Ising spins. Both these degeneracies are manifested themselves in the same residual entropy per elementary unit cell ${\cal S}=k_{\rm B}\ln 2$ and can be detected in the ground state of the model regardless of the nature (sign) of the spin-electron interaction. On the other hand, the field-independent two-fold degeneracy can be observed at $T=0$ only for the antiferromagnetic spin-electron interaction if triangular plaquettes are occupied by one or five mobile electrons and the total spin of the electron plaquette and the spin state of its Ising neighbor have opposite signs. A diversity of the ground-state degeneracy appearing in individual first-order phase transitions has been confirmed by low-temperature dependencies of the entropy. Finally, the effect of the ground-state degeneracy and the applied magnetic field on temperature dependence of the specific heat has been examined in detail. It has been evidenced that the mutual interplay of all factors may lead to complex temperature dependencies of the specific heat with two or three Schottky-type maxima.

\section*{Acknowledgments}
The work was financially supported by the grant No.~APVV-0097-12 of the Slovak Research and Development Agency under the contract and by the scientific grant VEGA 1/0043/16 of The Ministry of Education, Science, Research and Sport of the Slovak Republic.
Special thanks belongs to long-time collaborator Jozef Stre\v{c}ka for his valuable comments on this work.


\section*{References}
\begin{thebibliography}{20}
\bibitem{Bax82}
R.J. Baxter, Exactly Solved Models in Statistical Mechanics, Academic Press, New York, 1982.
\bibitem{Syo72}
I. Syozi, Phase Transition and Critical Phenomena, Academic Press, New York, 1972, pp. 269--329.
\bibitem{Per08}
M.S.S. Pereira, F.A.B.F de Moura, M.L. Lyra, Phys. Rev. B 77 (2008) 024402.
\bibitem{Nal14}
M. Nalbandyan, H. Lazaryan, O. Rojas, S.M. de Souza, N. Ananikian, J. Phys. Soc. Jpn. 83  (2014) 074001.
\bibitem{Gal15a}
L. G\'alisov\'a, J. Stre\v{c}ka, Phys. Rev. E 91 (2015) 022134.
\bibitem{Gal15b}
L. G\'alisov\'a, J. Stre\v{c}ka, Phys. Lett. A 379 (2015) 247.
\bibitem{Gal17}
L. G\'alisov\'a, D. Jakubczyk, Physica A 466 (2017) 30.
\bibitem{Str16}
J. Stre\v{c}ka, J. \v{C}is\'arov\'a, Mater. Res. Express 3 (2016) 106103.
\bibitem{Cis14}
J. \v{C}is\'arov\'a, J. Stre\v{c}ka, Phys. Lett. A 378 (2014) 2801.
\bibitem{Per09}
M.S.S. Pereira, F.A.B.F de Moura, M.L. Lyra, Phys. Rev. B 79 (2009) 054427.
\bibitem{Gal11}
L. G\'alisov\'a, J. Stre\v{c}ka, A. Tanaka, T. Verkholyak, J. Phys.: Condens. Matter 23 (2011) 175602.
\bibitem{Dor14}
F.F. Doria, M.S.S. Pereira, M.L. Lyra, J. Magn. Magn. Mater. 368 (2014) 98.
\bibitem{Has08}
M. Hase, H. Kitazawa, K. Ozawa, T. Hamasaki, H. Kuroe, T. Sekine,  J. Phys. Soc.  Jpn. 77 (2008) 034706.
\bibitem{Mat12}
M. Matsumoto, H. Kuroe, T. Sekine, M. Hase, J. Phys. Soc. Jpn. 81 (2012) 024711.
\bibitem{Der04}
O. Derzhko, J. Richter, Phys. Rev. B 770 (2004) 104415.
\bibitem{Hua63}
K. Huang, Statistical Mechanics, John Wiley and Sons, New York, 1963.
\end{thebibliography}
\end{document}